UDK 539.12

# Properties of a simple *e/γ* detector consisting of a lead convertor and a hodoscope

*S. P. Denisov, V. N. Goryachev, A. V. Kozelov*

Institute for High Energy Physics of the National Research Centre "Kurchatov Institute"
1 Nauki sqr, Protvino, Moscow region, 142281 Russian Federation

Vladimir.Goryachev@ihep.ru

**Abstract**

The results of the calculations of coordinate resolution and hadron rejection factor for a simple *e/γ* detector consisting of a lead converter followed by a hodoscope are presented. For the simulation of showers, initiated in the converter by electrons and hadrons with energies upto 1 TeV GEANT4 is used. It is shown that the best coordinate resolution for electrons is achieved when the converter thickness is closed to the position $t_{max}$ of the shower maximum. For example, at 200 GeV with 2 mm strip width hodoscope it is equal to $\sigma$=89 microns provided a "truncated mean" coordinate estimation is used. The optimal thickness of the converter for hadron rejection is also close to $t_{max}$. For 200 GeV beam of electrons and protons the rejection factor of $10^{-4}$ for 0.9 electron detection efficiency can be reached using only data on charged particles multiplicities. Information on the spatial distribution of the shower particles after the converter allows to enhance further the rejection by several times.

## 1. Introduction

Detectors consisting of a high *Z* converter and a hodoscope behind it were proposed by A.A. Tyapkin[1] as high energy *e/γ* spectrometers. They are widely used in experiments at accelerators and colliders for *e/γ* coordinate and energy measurement and *h/e* and *γ/π⁰* separation[2-20]. They are often referred to as shower maximum or preshower detectors. In this paper we present methods that can significantly improve their spatial resolution and enhance hadron rejection.

For the simulation of electromagnetic showers in the converter initiated by 10 to 1000 GeV electrons GEANT4 10.01.p02 (Physical list FTFP_BERT)[21] with 700 micron range cut for all particles is used. The corresponding energy thresholds in lead are 1 MeV for $e^+$ and $e^-$ and 0.1 MeV for the *γ*. Increase or decrease of the range cut by factor of 2 does not change $e^+$ and $e^-$ multiplicities in showers within statistical error of 0.5%[18]. The same GEANT4 version is used to simulate the passage of protons through the converter.

The results below are for the lead converter unless otherwise specified. The diameter of the converter was chosen to be equal to 70 cm. The thickness of the converter *t* is measured in radiation length units $X_0$ and the electron energy $E_0$ is in GeV. The thickness $t_{max}$ corresponding to the maximum flux of the shower particles as a function of the electron energy $E_0$ is described by the formula[18]:

$$t_{max} = 1.11 \ln E_0 + 3.14.$$

For frequently used energies of 40, 80, 200, and 500 GeV $t_{max}$ is equal to 7.2, 8.0, 9.0, and 10.0 $X_0$. It is assumed that the trajectory of the primary electrons is perpendicular to the hodoscope plane. The root-mean-square deviations of statistical distributions are denoted below as RMS or $\sigma$.

## 2. Coordinate resolution

The coordinate resolution of the $e/\gamma$ detector depends on the spatial distribution of charged particles after the converter and the hodoscope structure. Integral radial distributions of particles at the shower maximum are presented in ref./19/. In the range of $r$ up to ~20 g/cm$^2$, containing about 98% of particles, they are reasonable well fitted by a sum of two exponents:

$$N(r)/N_0 = 1 - f_0 e^{-sr} - (1-f_0) e^{-tr}, \qquad (1)$$

where $N_0$ is the total number of particles and $f_0$, $s$, $t$ are free parameters that weakly depend on the energy and Z of the converter, if $r$ is expressed in g/cm$^2$ (see Fig. 1 and/19/). Below differential distributions of particles along the transverse coordinate $x$ are used. If the radial distribution satisfies the equation (1), $x$ distribution is described by the sum of two cylindrical $K_0$-functions:

$$\frac{1}{N_0} \cdot \frac{dN}{dx} = \frac{1}{\pi}\left[ s f_0 K_0(sx) + t(1-f_0) K_0(tx) \right] \qquad (2)$$

An example of such distribution is shown in Fig. 2. In the region from -10 to +10 g/cm$^2$, containing more than 96% of particles, it is well described by formula (2). Note that the differential distributions are rather narrow (in the distribution on Fig. 2 80% of the particles are in the range from -2.2 to 2.2 mm) but they have long "tails".

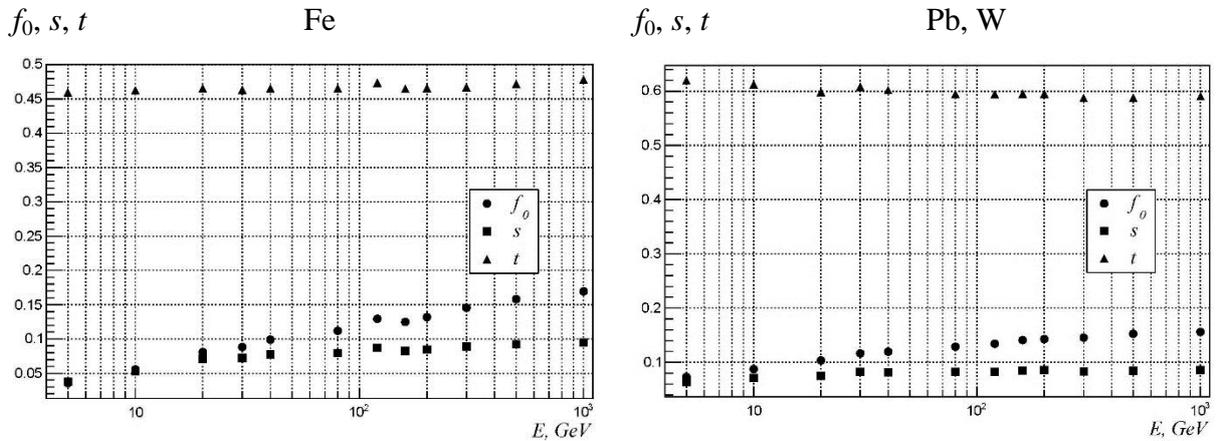

Fig. 1. The dependence of the parameters in formula (1) on the electron energy. The difference in parameter values for Pb and W is less than the marks size.

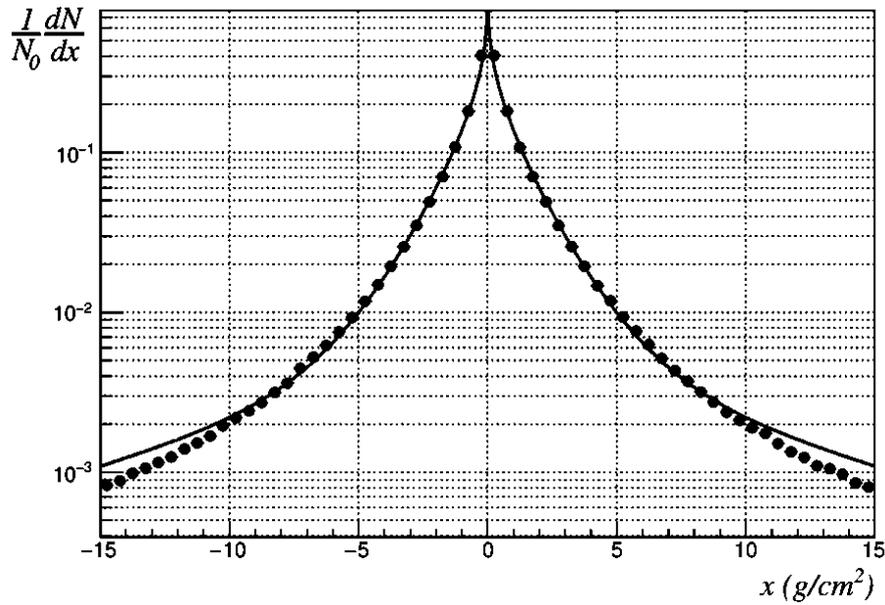

Fig. 2. The distribution of charged particles along the transverse coordinate after the converter of $9X_0$ for $E_0 = 200$ GeV. Statistical errors do not exceed the marks size. The curve represents equation (2) with the parameters $f_0=0.14$, $s = 0.086$, $t = 0.59$ from ref./19/.

A detector consisting of strips with width $d$ is considered as a hodoscope. In such a detector a shower axis coordinate $\bar{x}$ is often estimated by the center of gravity method using information on the number of particles (signal amplitude) in all strips. However, the uncertainty of a such estimate can be significantly reduced for some distributions if so-called "truncated mean"/22/ $\bar{x}_\alpha$ instead of $\bar{x}$ is used. Only central strips containing $N_i/2-\alpha$ particles on each side of $\bar{x}$ are participating in the $\bar{x}_\alpha$ calculation by this method, where $N_i$ is a total number of particles and $\alpha$ is a fraction of ignored peripheral particles on each side. The optimal value of $\alpha$, minimizing the coordinate resolution, depends on $d$. This dependence is illustrated in Table 1. Sample distributions of $\bar{x}$ and $\bar{x}_\alpha$ uncertainties for $\alpha$ values close to the optimal are shown in Fig. 3. The method of "truncated mean" is effective if the strip size $d$ is comparable or less than the half maximum width of the particle distribution (for $E_0=200$ GeV and $9X_0$ converter thickness it is equal to 3.6 mm). For $d=1$ and 2 mm it allows to improve the resolution by factors of 5 and 3 (see Fig. 3 and Table 1), while for $d>4$ mm there is no substantial improvement.

Table 1. Dependence of RMS (μm) on $\alpha$ for the $9X_0$ converter and $E_0=200$ GeV.

| $\alpha$ | 0 | 0.02 | 0.04 | 0.06 | 0.08 | 0.10 | 0.12 | 0.14 | 0.16 | 0.18 | 0.20 | 0.22 | 0.24 | 0.26 | 0.28 |
|---|---|---|---|---|---|---|---|---|---|---|---|---|---|---|---|
| $d=1$ мм | 359 | 118 | 97 | 91 | 81 | 78 | 77 | 77 | 74 | 71 | 68 | 69 | 70 | 74 | 86 |
| $d=2$ мм | 397 | 149 | 138 | 135 | 138 | 133 | 129 | 135 | 149 | 173 | 206 | 234 | 265 | 296 | 327 |
| $d=4$ мм | 508 | 378 | 388 | 386 | 412 | 457 | 535 | 580 | 640 | 688 | 734 | 779 | 823 | 861 | 878 |

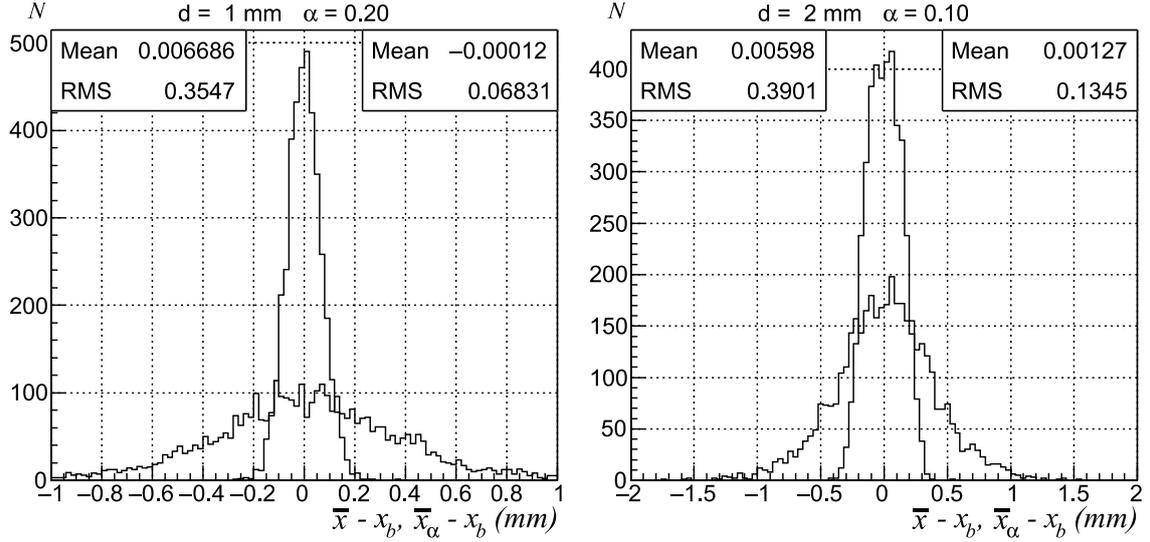

Fig. 3. Uncertainties distributions of the $\bar{x}$ and $\bar{x}_\alpha$ estimates of the shower axis for $E_0=200$ GeV and $9X_0$ thick converter: $\bar{x}$---wide histograms and $\bar{x}_\alpha$---narrow histograms, $x_b$---true coordinate of the shower axis. Values of $\alpha$ are close to optimal (see Table 1).

It's known that the center of gravity method leads to biased estimate of the shower axis coordinate, if the trajectory of the primary particles does not pass through the center or the edge of a strip (see., e.g.,/23/). To find an uncertainty of $\bar{x}_\alpha$ due to this effect, the normalized values $\hat{x}_\alpha = (\bar{x}_\alpha - x_0)/d$ and $\hat{x}_b = (x_b - x_0)/d$ where calculated for events uniformly distributed across one of the strips ($x_0$ is a coordinate of the center of the strip containing $\bar{x}_\alpha$, and $x_b$ is $x$ coordinate of the primary electron). Fig. 4 shows dependencies of $\hat{x}_b$ vs $\hat{x}_\alpha$ for different $d$ approximated by the modified logistic function

$$f(S) = \frac{1+e^{-A}}{1-e^{-A}} \cdot \left( \frac{1}{1+e^{-A \cdot S}} - \frac{1}{2} \right), \tag{3}$$

where $S(x) = \sum_{i=0}^{k} a_{2i+1} \cdot T_{2i+1}(2x)$ is the sum of Chebyshev polynomials of the first kind, and $A$ and $a_{2i+1}$ are free parameters. The condition $a_1 = 1 - \sum_{i=1}^{k} a_{2i+1}$ is imposed on the parameter $a_1$ which ensures the equality $\hat{x}_b = \hat{x}_\alpha = \pm 0.5$ at the strip ends. Equality $\hat{x}_b = \hat{x}_\alpha = 0$ at the strip center is automatically satisfied by using the odd Chebyshev polynomials. The initial value of $k$ is chosen to be 11. Then starting with the highest degree, the significance of the coefficients $a_{2i+1}$ is checked. If the absolute value of the parameter is less than its tripled error, the value of $k$ is decreased by one, and fit is repeated with fewer number of parameters. The final value of $k$ depends on $d$ and, for example, for a converter of $9X_0$ varies from 0 ($d=1$ mm) to 6 ($d=16$ mm). The dependence $\hat{x}_b(\hat{x}_\alpha)$ can be described by the Chebyshev polynomials only but using the logistic function allows to reduce the number of free parameters.

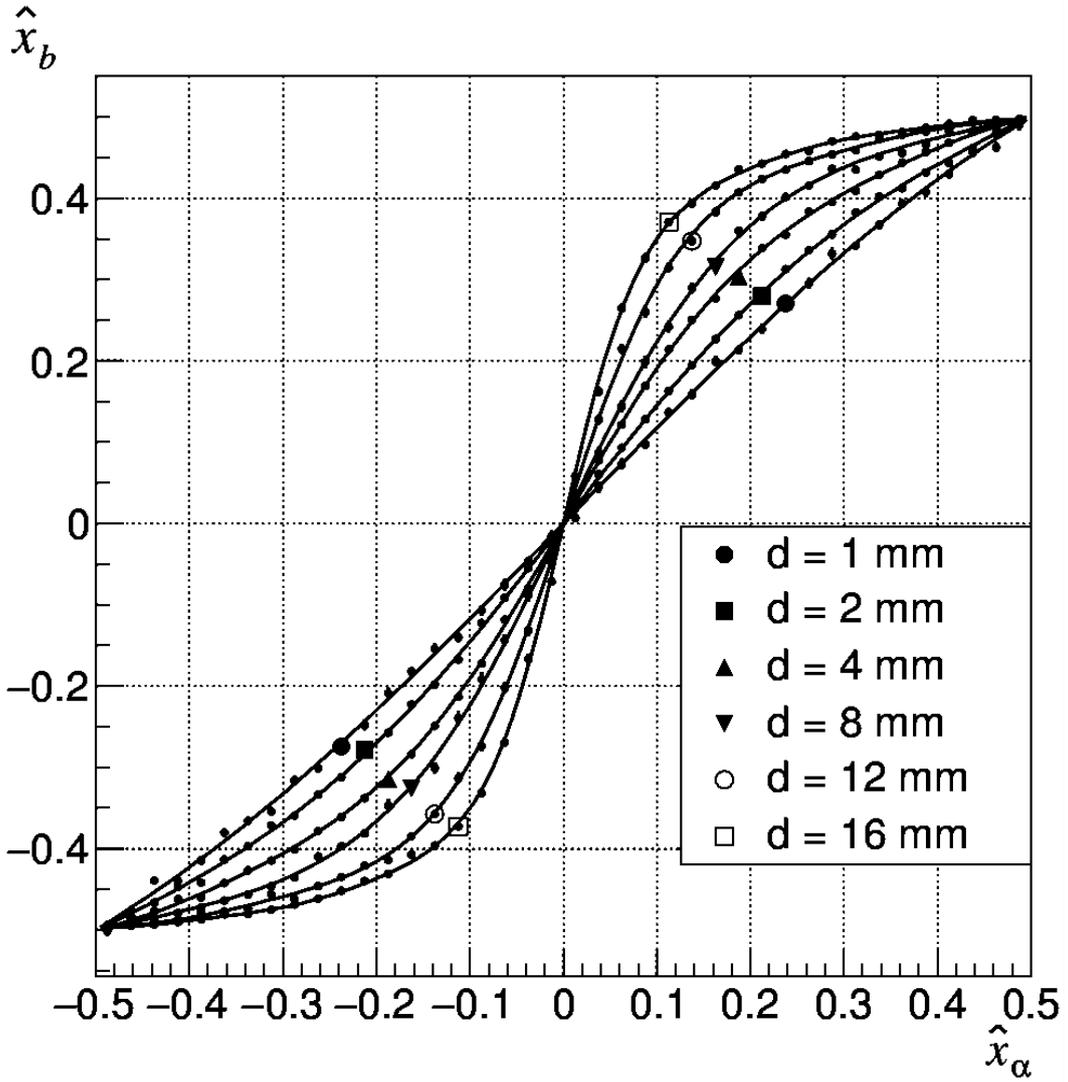

Fig. 4. The relationship between the reconstructed $\hat{x}_\alpha$ and the true $\hat{x}_b$ coordinates of the shower axis for $E_0$=200 ГэВ and $9X_0$ thick converter: $\hat{x}_\alpha = (\bar{x}_\alpha - x_0)/d$ and $\hat{x}_b = (x_b - x_0)/d$, where $x_0$---coordinate of the center of strip containing $\bar{x}_\alpha$. The simulation results are fitted to function (3). The $\chi^2/ndf$ values are close to 1 for all curves.

The proposed method for correction of the $\bar{x}_\alpha$ bias is tested with the statistics not used to determine the function $f$ parameters. Fig. 5 presents the distributions of $x_c$ obtained by applying function $f(x)$ for the bias correction of $\bar{x}_\alpha$. Comparison of Fig. 3 and Fig. 5 shows that the bias correction is more important for the wide strips. It allows, for example, to reduce RMS by a factor of 2.5 for 4 mm strip, while for 1 mm strip there is almost no improvement in resolution.

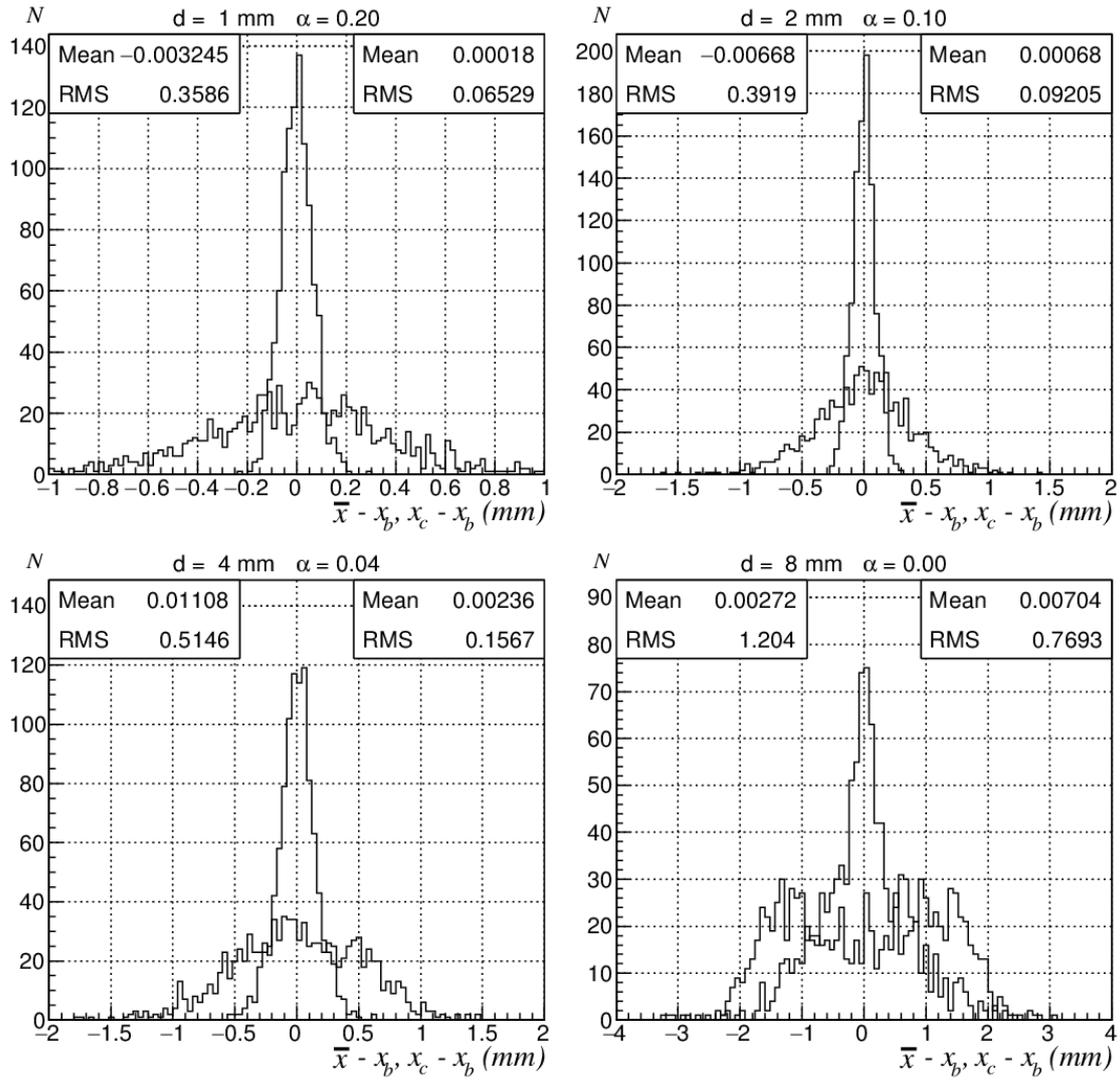

Fig. 5. Distributions of uncertainties of $\bar{x}$ and $x_c$ estimates for $E_0$=200 GeV and $9X_0$ thick converter ($x_c$ ---coordinate of the shower axis reconstructed by the "truncated average" method with bias correction); $\bar{x}$---wide histograms, $x_c$ ---narrow histograms, $x_b$---true coordinate of the shower axis.

Fig. 6--8 illustrate the dependence of coordinate resolution of $e/\gamma$ detector on the converter thickness $t$, shower energy $E_0$ and the hodoscope strip width $d$. From Fig. 6--7 it follows that in the region of the shower maximum the function $x_c(t)$ passes through a wide minimum, which is consistent with the measurements of ref./24/, reaching, for example, at $E_0$=200 GeV, $t=9X_0$ and $d=2$ mm the value of 89 μm. Slight difference in RMS values in Fig. 5 ($d=2$ mm) and Fig. 6 ($t=9$) is associated with the use of different statistical samples. The dependencies of the $x_c$ uncertainties on $E_0$ for 1, 2, and 4 mm strips shown in Fig. 8 are fitted to the function

$$\sigma(x_c - x_b) = A + B/\sqrt{E_0}. \qquad (4)$$

The values of parameters $A$ and $B$ are given in Table 2.

Table 2. The values of the parameters in formula (4).

| $d$, мм | 1 | 2 | 4 | 8 |
|---|---|---|---|---|
| A | 0.010 ± 0.001 | 0.012 ± 0.001 | 0.029 ± 0.003 | 0.283 ± 0.013 |
| B | 0.758 ± 0.011 | 1.084 ± 0.018 | 1.776 ± 0.033 | 6.16 ± 0.15 |

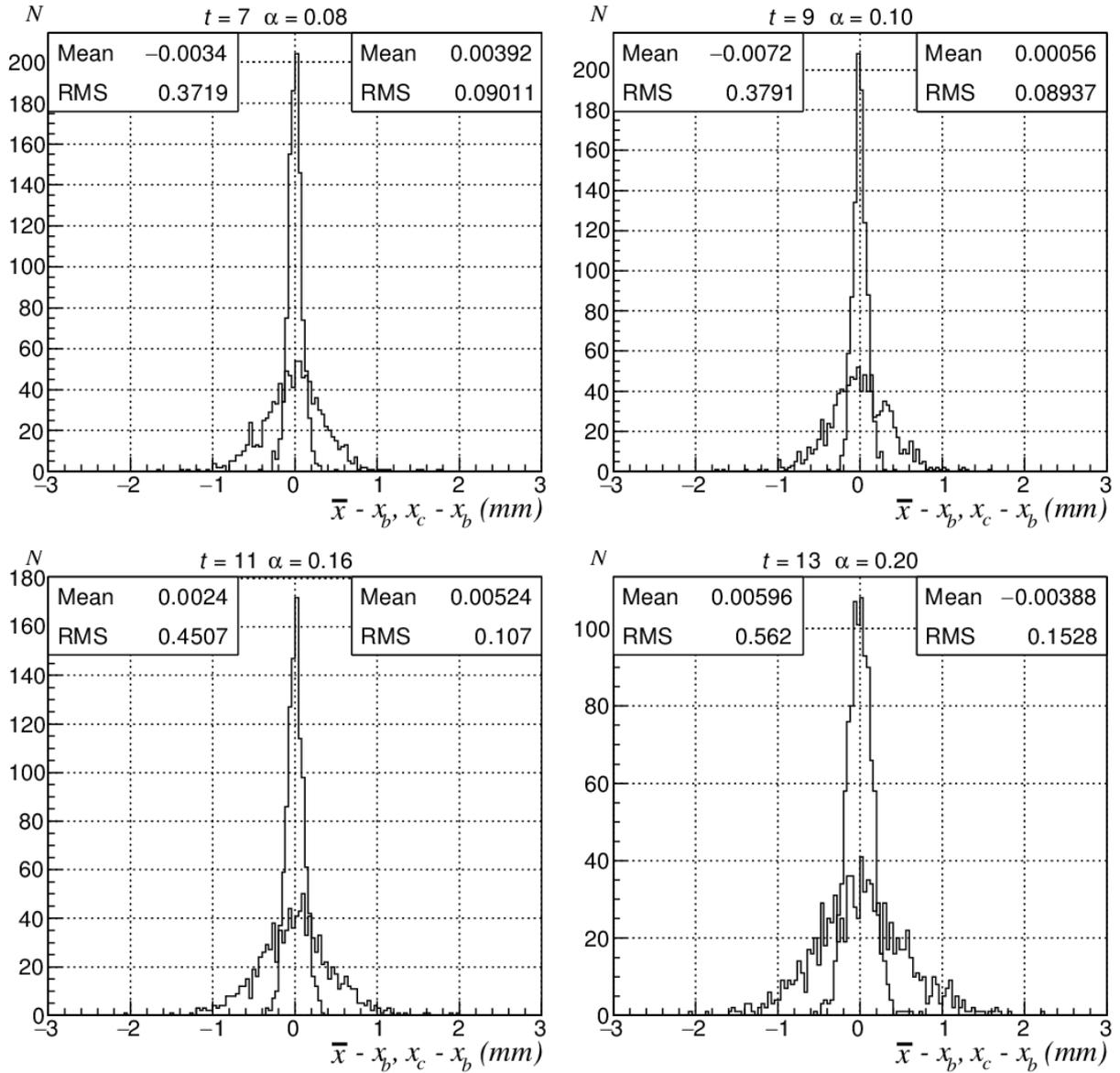

Fig. 6. Distributions of the uncertainties of $\bar{x}$ and $x_c$ for $E_0$=200 GeV and for 2 mm strips; $\bar{x}$---wide histograms, $x_c$---narrow histograms, $x_b$---true coordinate of the shower axis. The values of $\alpha$ shown are close to optimal.

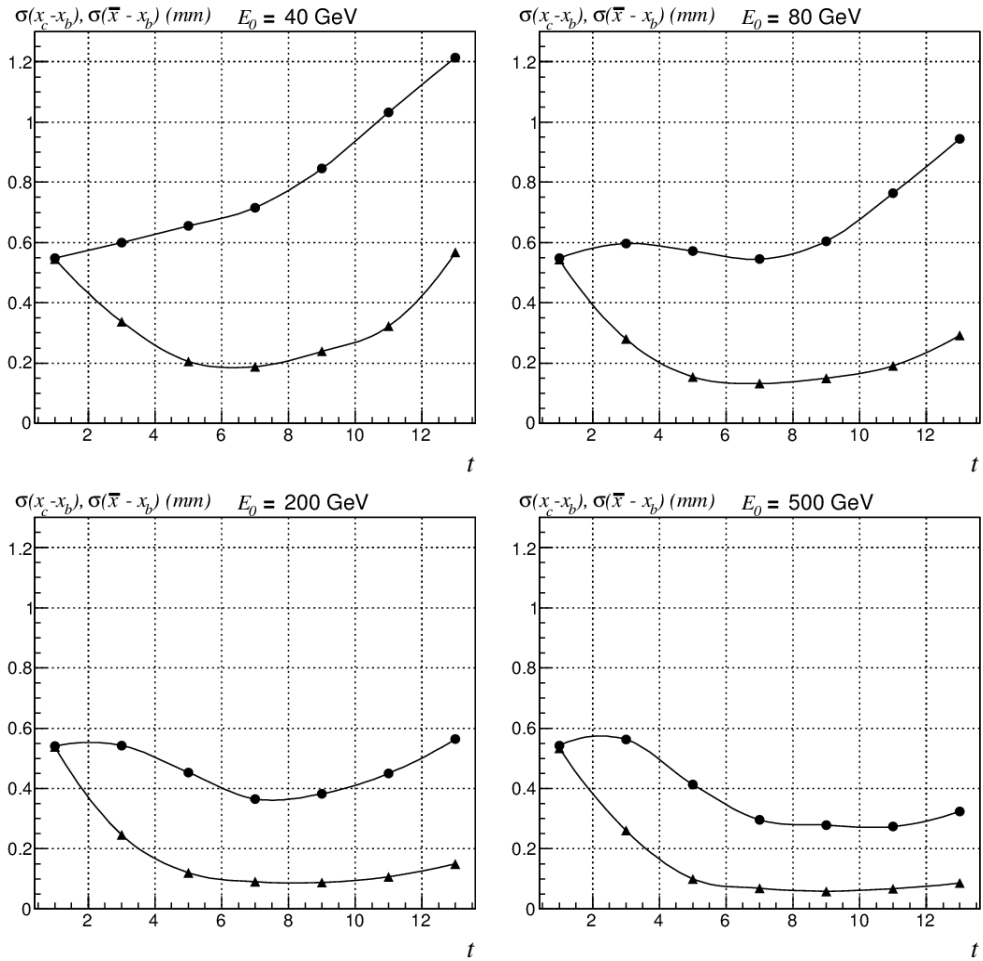

Fig. 7. The dependencies of the $\bar{x}$ and $x_c$ uncertainties on the converter thickness for different shower energies and 2 mm strips. The curves are drawn using ROOT.

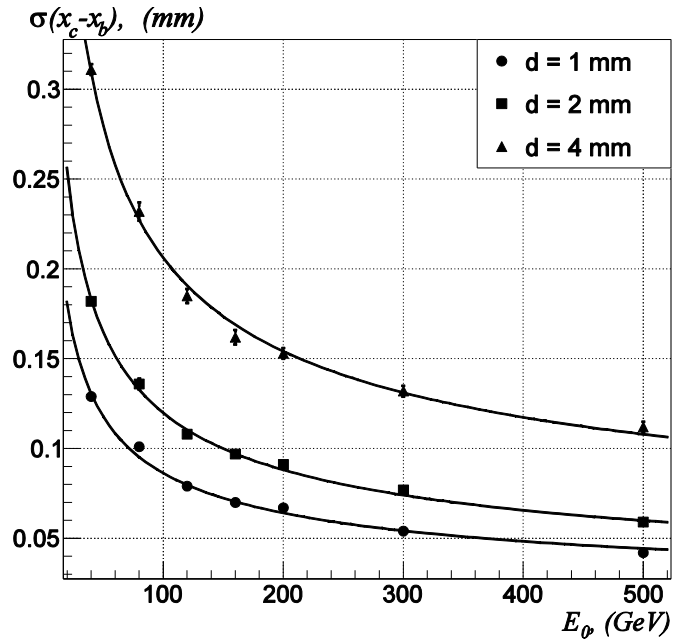

Fig. 8. The dependencies of $x_c$ uncertainties vs shower energy for the $\alpha$ values closed to optimal; $x_b$---true coordinate of the shower axis, $d$---strip width. Results are fitted to formula (4) with parameters shown in Table 2. Errors are close to the marks size.

## 3. Hadron rejection

To determine the hadron rejection factor $h/e$, multiplicity distributions of charged particles for protons and electrons with energies of 40, 80, 200 and 500 GeV are calculated (examples of such distributions are presented in Fig. 9). Using the distributions for the electrons, the multiplicity values corresponding to the electron detection efficiency $\varepsilon_e$ 0.90, 0.95 and 0.99 (shown by arrows on Fig. 9) are determined.

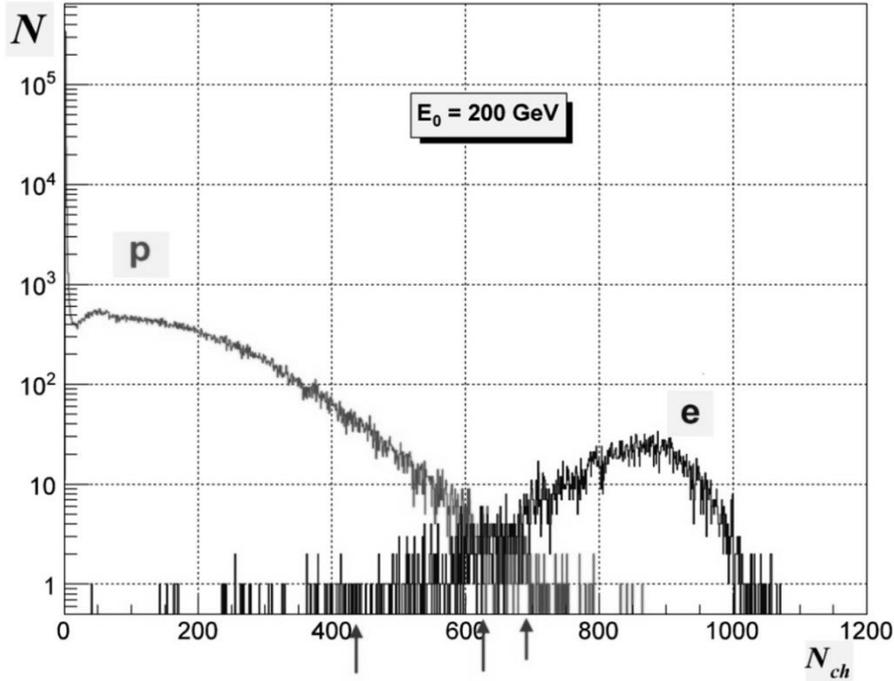

Fig. 9. Distributions of the charged particles multiplicities for the proton and electron showers. The arrows indicate the multiplicity values corresponding to the electron detection efficiency of 0.99, 0.95, and 0.90 (from left to right).

The protons detection efficiency $\varepsilon_p$ for the given $\varepsilon_e$ value is determined as the ratio of the number of hadron events with a multiplicity greater than the threshold for electrons to the total number of proton events. The obtained dependencies of $\varepsilon_p$ on the converter thickness $t$ are shown on Fig. 10.

All dependencies have a minimum at the converter thickness of $t_{min}$ which is close to $t_{max}$ in agreement with experimental data/4/. The differences between $t_{min}$ and $t_{max}$ are approximately 0.5, 1.0, and 1.5 $X_0$ for $\varepsilon_e$ 0.90, 0.95 and 0.99 independent of energy. This means that $t_{min}$ weakly (logarithmically as $t_{max}$) depends on $E_0$ that allows to achieve a very low ($10^{-3}$--$10^{-4}$) proton detection efficiency in a wide energy range with the same converter thickness. One can expect that for mesons $\varepsilon_m < \varepsilon_p$ since their free path to inelastic interaction in lead is 1.2 times greater than those for nucleons/25/. The obtained $\varepsilon_p$ values do not take into account the amplitude resolution of the hodoscope and therefore are the lowest estimates. For example, in the experiment /4/ performed in the 40 GeV/c beam $\varepsilon_p=4\cdot 10^{-3}$ was obtained for $\varepsilon_e=0.95$ and $t_{max}$ converter thickness.

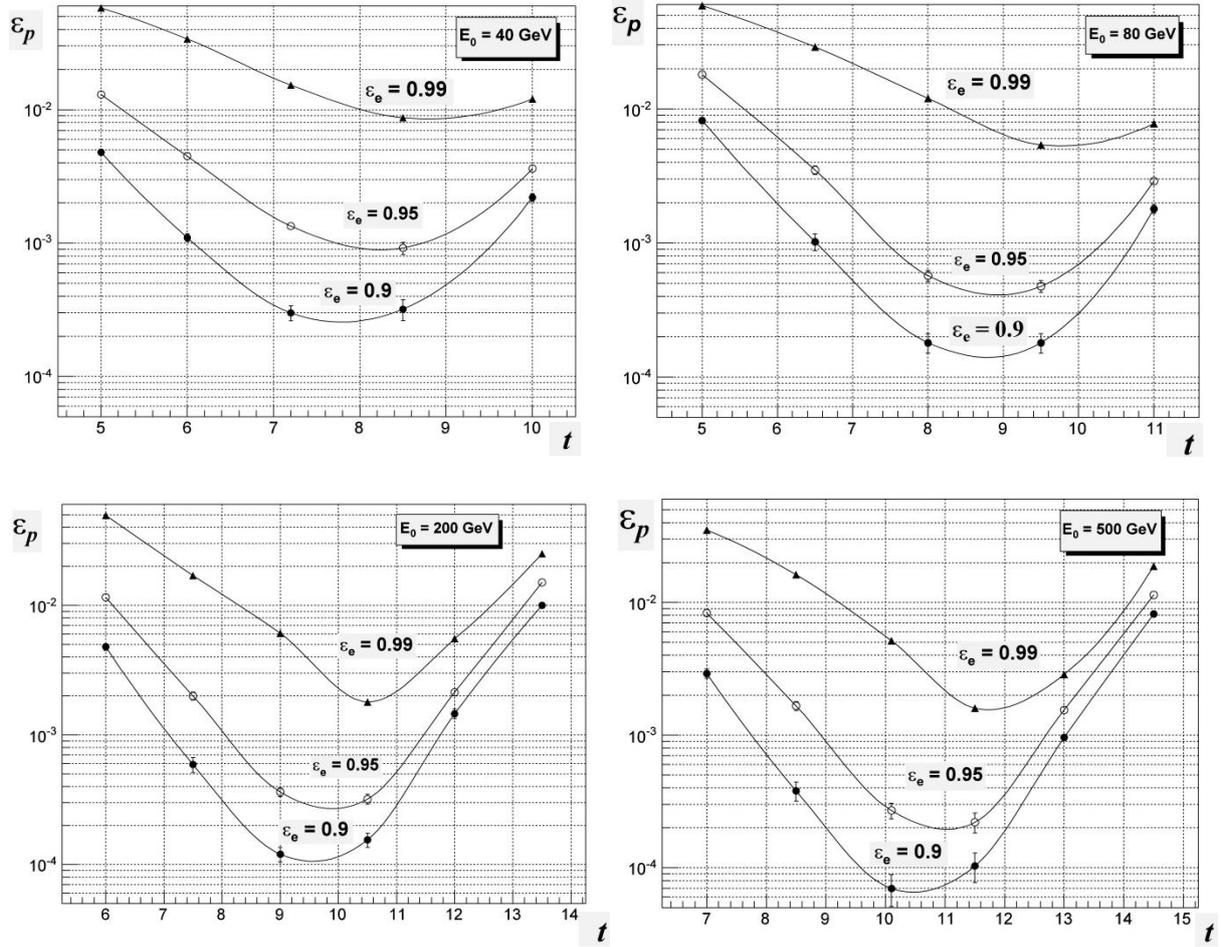

Fig. 10. Dependencies of the proton detection efficiency on the converter thickness.

In addition to the multiplicity the difference in the spatial distributions of the charge shower particles, which are wider for hadron showers, can also be used for proton rejection. We have studied the possibility to enhance the rejection by introducing restrictions on the distance $r$ of the detected particles from the shower axis. In the Table 3 $\varepsilon_p$ values are given for particles with $r$ values less than 1, 2, 5, 10, and 350 mm for $E_0=200$ GeV and a converter thickness of $9X_0$ (350 mm is the converter radius). The fractions of the particles in the electromagnetic showers inside these rings are 0.44, 0.66, 0.88, 0.95, and 1.00 /19/. Table 3 shows that the optimum $r_{max}$ value depends on $\varepsilon_e$ and significant (by a factor of 3) decrease of $\varepsilon_p$ can be obtained for $\varepsilon_e=0.99$, while for $\varepsilon_e=0.90$ the effect is 25% only.

Table 3. The dependence of the proton detection efficiency on $r_{max}$ for $E_0 = 200$ GeV and a converter thickness of $9X_0$. 350 mm is the converter radius.

| $r_{max}$, mm | $\varepsilon_p$ | | |
|---|---|---|---|
| | $\varepsilon_e=0.90$ | $\varepsilon_e=0.95$ | $\varepsilon_e=0.99$ |
| 1 | $1.8 \cdot 10^{-4}$ | $3.4 \cdot 10^{-4}$ | $2.1 \cdot 10^{-3}$ |
| 2 | $8.4 \cdot 10^{-5}$ | $2.1 \cdot 10^{-4}$ | $2.7 \cdot 10^{-3}$ |
| 5 | $7.3 \cdot 10^{-5}$ | $2.2 \cdot 10^{-4}$ | $3.9 \cdot 10^{-3}$ |
| 10 | $8.0 \cdot 10^{-5}$ | $3.2 \cdot 10^{-4}$ | $4.7 \cdot 10^{-3}$ |
| 350 | $9.6 \cdot 10^{-5}$ | $3.8 \cdot 10^{-4}$ | $6.4 \cdot 10^{-3}$ |

Besides $r$ another parameter could be the RMS of the transverse shower profile. An example of a RMS probability density distribution is shown in Fig. 11. For electrons all events are used, for protons only those with multiplicity above the threshold for $\varepsilon_e=0.99$. Data analysis has shown that, for example, with RMS cuts of 5.3 and 4.9 mm it is possible to get further improvement in hadron rejection by factors of 2 ($\varepsilon_e=0.95$) and 3 ($\varepsilon_e=0.90$). The above estimates of the hadron rejection factor do not take into account the properties of a shower particle detector, for example, its spatial resolution.

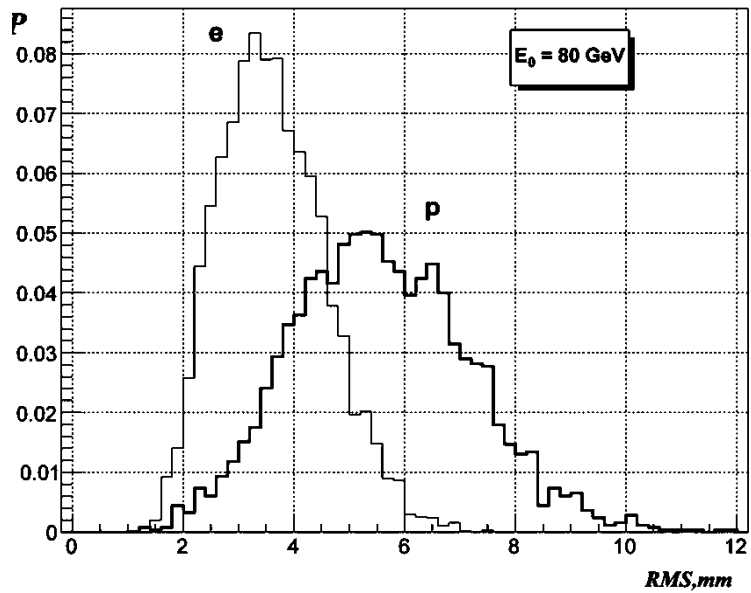

Fig. 11. Normalized *RMS* distributions for protons and electrons for a converter of thickness of $t_{max}=8X_0$. Only events with multiplicity above the threshold corresponding to the electron detection efficiency 0.99 are used for protons.

## 4. Conclusion

Different methods that allow to improve the characteristics of the *e/γ* detector consisting of a lead converter followed by a hodoscope are considered. Simulation of the showers initiated in a converter by electrons and protons are performed using GEANT4. It is shown that precision of electron coordinate reconstruction can be improved by a factor from 3 to 5 if a "truncated mean" instead of a conventional center of gravity method is used and Chebyshev polynomials are applied to compensate the systematic bias associated with the finite size of the hodoscope elements. In particular, for the shower energy of 200 GeV with the hodoscope strip width of 2 mm, the proposed technique allows to achieve a resolution of 89 μm. Another important characteristic of the $e/\gamma$ detector is hadron rejection factor. It turned out that the best hadron rejection as well as the best coordinate resolution are achieved with a converter thickness close to the position of the shower maximum. For example, for a 200 GeV beam of electrons and protons and $t=9X_0$ the rejection factors of $4\cdot10^{-4}$ and $6\cdot10^{-3}$ for 0.95 and 0.99 electron detection efficiency can be achieved, if only data on multiplicity of shower charged particles are used. Information about spatial distribution of charged particles allows to enhance further the hadron rejection several times. Thus, the coordinate resolution and hadron rejection factor of the considered detector are close to similar characteristics of the complex and expensive electromagnetic calorimeters.

The authors are gratefully acknowledge the help of D.S. Denisov, T.Z. Gurova and D.A. Stoyanova in preparation of this manuscript. This work was supported in part by the Russian Foundation for Basic Research under grant #17-02-00120.

**References**

[1] *Tyapkin A.A.* // NIM, 1970. V. 85. P. 277-278.

[2] *Muller D.* // Phys. Rev. D, 1972. V. 5. P. 2677.

[3] *Amatuni Ts.A., Denisov S.P., Krasnokutsky R.N. et al*. // NIM, 1982. V. 203. P. 179-182.

[4] *Amatuni Ts.A., Antipov Yu.M., Denisov S.P. et al*. // NIM, 1982. V. 203. P. 183-187.

[5] *Zhang C.S., Shibata M., Kasahara K., Yuda T.* // NIM A, 1989. V. 283. P. 78-87.

[6] *del Peso J., Ros E.* // NIM A, 1991. V. 306. P. 485.

[7] *Apollinari G., Giokaris N.D., Goulianos K. et al*. // NIM A, 1993. V. 324. P. 475-481.

[8] *Acosta D., Bylsma B., Durkin L.S. et al*. // NIM A, 1995. V. 354. P. 296-308.

[9] *Alvsvaag S.J., Maeland O.A., Klovning A. et al*. // NIM A, 1995. V. 360. P. 219-223.

[10] *Byrum K., Dawson J., Nodulman L. et al*. // NIM A, 1995. V. 364. P.144-149.

[11] *Akimenko S.A., Belousov V.I., Chujko B.V. et al*. // NIM A, 1995. V. 365. P. 92-97.

[12] *Grunhaus J., Kananov S., Milststene C.* // NIM A, 1995. V. 354. P. 368-375.

[13] *Chang Y.H., Chen A.E., Hou S.R. et al*. // NIM A, 1997. V. 388. P. 135-143.

[14] *Kawagoe K., Sugimoto Y., Takeuchi A. et al*. // NIM A, 2002. V. 487. P. 275-290.

[15] *Balanda A., Jaskula M., Kajetanowicz M. et al*. // NIM A, 2004. V. 531. P. 445-458.

[16] *Itoh S., Takeshita T., Fujii Y., Kajino F. et al*. // NIM A, 2008. V. 589. P. 370-382.

[17] *Ronzhin A., Los S., Ramberg E. et al.* // NIM A, 2015. V. 795. P. 288-292.

[18] *Denisov S.P., Goryachev V.N.* // Physics of Atomic Nuclei, 2018. V. 81. № 10. P. 1488-1493.

[19] *Denisov S.P., Goryachev V.N.* // arXiv:1812.07906 [physics.ins-det], 2018.

[20] *Denisov S.P., Goryachev V.N.* // arXiv:1812.10054 [physics.ins-det], 2018.

[21] http://cern.ch/geant4.

[22] *Leman E.* // Moscow. "Nauka", 1991. *E.L. Lehmann* // John Wiley and sons, 1983.

[23] *Akopdjanov G.A., Inyakin A.V., Kachanov V.A. et al.* // NIM A, 1977. V. 140. P. 441-445.

[24] *Amatuni Ts.A., Antipov Yu.M., Denisov S.P., Petrukhin A.I.* // ПТЭ, 1983. V. 3. P. 33.

[25] *Gorin Yu.P., Denisov S.P., Donskov S.V. et al.* // ЯaF, 1973. V. 18. 336. // Nucl. Phys B, 1973. V. 61. P. 62.